\documentclass[twocolumn]{IEEEtran}

\usepackage{xspace,amsmath,amssymb,amsfonts,epsfig,syntonly}
\usepackage{cite,bm,color,url,textcomp,empheq,boxedminipage}
\usepackage{algorithmicx,algorithm,algpseudocode}
\usepackage{epstopdf}
\usepackage{empheq}
\usepackage{pifont}

\usepackage{graphicx,graphics}  
\usepackage{multirow,multicol}
\usepackage{psfrag}    
\usepackage{stfloats}
\usepackage{url}

\newtheorem{lemma}{\underline{Lemma}}[section]

\newcommand{\mv}[1]{\mbox{\boldmath{$ #1 $}}}

\long\def\symbolfootnote[#1]#2{\begingroup
\def\thefootnote{\fnsymbol{footnote}}
\footnote[#1]{#2}\endgroup}

\psfull

\allowdisplaybreaks[4]

\usepackage[T1]{fontenc}
\usepackage{ifpdf}
 \ifpdf
 \else
 \fi

\begin{document}
\title{Wireless Powered User Cooperative Computation in Mobile Edge Computing Systems}
\author{Dixiao Wu, Feng Wang, Xiaowen Cao, and Jie Xu\\
School of Information Engineering, Guangdong University of Technology, Guangzhou, China\\
Email: 975149733@qq.com, fengwang13@gdut.edu.cn, caoxwen@outlook.com, jiexu@gdut.edu.cn
\thanks{$^*$J. Xu is the corresponding author. This work was supported in part by the National Natural Science Foundation of China (Project No. 61871137).}
}

\maketitle


\begin{abstract}
This paper studies a wireless powered mobile edge computing (MEC) system, where a dedicated energy transmitter (ET) uses the radio-frequency (RF) signal enabled wireless power transfer (WPT) to charge wireless devices for sustainable computation. In such a system, we present a new user cooperation approach to improve the computation performance of active devices, in which surrounding idle devices are enabled as helpers to use their opportunistically harvested wireless energy from the ET to help remotely execute active users' computation tasks. In particular, we consider a basic scenario with one user (with computation tasks to execute) and multiple helpers, in which the user can partition the computation tasks into various parts for local execution and computation offloading to helpers, respectively. Both the user and helpers are subject to the so-called energy neutrality constraints, such that their energy consumption does not exceed the respective energy harvested from the ET. Under this setup and considering a frequency division multiple access (FDMA) based computation offloading protocol, we maximize the computation rate (i.e., the number of computation bits over a particular time block) of the user, by jointly optimizing the transmit energy beamforming at the ET, as well as the communication and computation resource allocations at both the user and helpers. By leveraging the Lagrange duality method, we present the optimal solution to this problem in a semi-closed form. Numerical results show that the proposed wireless powered user cooperative computation design significantly improves the computation rate at the user, as compared to conventional schemes without such cooperation.
\end{abstract}


\section{Introduction}
With recent advancements in artificial intelligence, big data, and internet of things (IoT), it is envisioned that future wireless networks need to support massive low-power wireless devices (e.g., sensors and actuators) with real-time communication and computation, in order to enable various new applications such as industrial automation, smart transportation, and unmanned aerial vehicles (UAVs). Towards this end, how to provide rich computation capability and sustainable energy supply for these wireless devices is becoming a critical technical challenge to be tackled.

Recently, mobile edge computing (MEC) has emerged as a promising solution to enhance wireless devices' computation capability\cite{1,2,3}. Different from conventional mobile cloud computing (MCC) with centralized clouds that are normally far apart from wireless devices, MEC offers remote computation services at the network edge in their close proximity. By allowing wireless devices to offload computation tasks to nearby base stations (BSs), WiFi access points (APs), or even smart phones and laptops for remote execution in MEC, these devices can enjoy enhanced computation capability and reduced computation latency. On the other hand, radio-frequency (RF) signal based wireless power transfer (WPT) has been recognized as a viable and convenient solution to charge low-power electronic devices by deploying dedicated energy transmitters (ETs) for energy broadcasting (see, e.g., \cite{4,5,6,7}). Simultaneous wireless information and power transfer (SWIPT) and wireless powered communication networks (WPCNs) are two main WPT applications that aim to provide sustainable wireless communications in the IoT era \cite{8,9,10,11}.

To exploit both benefits of MEC and WPT, wireless powered MEC has been recently proposed to achieve self-sustainable computing for wireless devices, in which a new type of hybrid APs are deployed to not only serve as ETs to wirelessly charge devices, but also act as edge servers to help remotely execute their offloaded computation tasks\cite{12,13,14,15,APCC17}. The work in \cite{12} first considered a wireless powered single-user MEC system, in which the WPT at the AP as well as the local computing and computation offloading at the users are jointly optimized, to maximize the user's successful computation probability, subject to the computation latency constraints. The work in \cite{13} further investigated a wireless powered multiuser MEC system under a time-division multiple access (TDMA) based partial offloading protocol, in which the multi-antenna energy beamforming at the AP and the computation/offloading decisions at the users are jointly optimized to minimize the overall energy consumption of the system, subject to the users' computation latency constraints. Furthermore, the authors in \cite{14,APCC17} and \cite{15} studied a wireless powered multiuser MEC system with computation rate maximization and a wireless powered single-relay system for MEC, respectively.

Despite the research efforts on wireless powered MEC, the above works \cite{12,13,14,15,APCC17} focused on the scenario with a centralized edge server co-located at the ET (i.e., the hybrid AP). Such a design, however, is generally not applicable in other WPT scenarios when ETs are dedicatedly deployed without computation capabilities. Also, this design fails to exploit the rich computation resources at surrounding end users. It is worth noting that nowadays, smart IoT devices are densely deployed in wireless networks. Due to the burst nature of wireless traffic, it is highly likely that, when some devices are actively computing, there exist some surrounding idle devices with unused computation resources. Thanks to the broadcast characteristics of WPT, these idle devices can also efficiently harvest wireless energy from ETs. Motivated by these facts, we propose a new wireless powered user cooperative computing approach to exploit both the unused computation resources and the opportunistic wireless energy harvesting at surrounding idle devices, in which these devices are enabled as helpers to use their opportunistically harvested wireless energy to help remotely execute the active users' computation tasks, thus improving the computation performance. Notice that the cooperative computation between two users has been investigated in our previous work\cite{16}, the energy-efficient multiuser computation offloading designs based on a non-orthogonal multiple access (NOMA) protocol have been pursued for improving the MEC system performance\cite{NOMA}, and the so-called federated learning has been developed by Google to enable multiple mobile phones to collaborate in executing machine learning tasks\cite{17}. However, these works only considered the users' cooperative computation under fixed energy supplies (e.g., batteries), while our work in this paper unifies both cooperative computing and WPT, where the energy consumption of helpers comes from the wireless energy transferred from the ET, thus leading to self-sustainable computation cooperation among users.

In this paper, we consider a wireless powered multiuser MEC system consisting of a multi-antenna ET, an active-computing user, and multiple helpers for cooperative computing. The ET broadcasts wireless energy to charge all the user and helpers simultaneously. Relying on the harvested energy, the user can partition its computation tasks into various parts that are computed locally and offloaded to multiple helpers for parallel execution, respectively. In order to avoid the co-channel interference, we consider that the WPT and the computation task offloading are implemented over orthogonal frequency bands. Furthermore, a frequency-division multiple access (FDMA) protocol is adopted for the task offloading and result downloading between the user and different helpers. For the cooperative computation between the user and each helper, the computation time block of our interest is divided into three time slots, for task offloading from the user to the helper, the helper's task execution, and the computation results downloading from the helper to the user, respectively. Under this setup, we maximize the computation rate (i.e., the number of computation bits over a particular time block) at the user, by jointly optimizing the transmit energy beamforming at the ET, as well as the communication and computation resource allocations at both the user and helpers, subject to their energy neutrality constraints (i.e., their energy consumption does not exceed the respective energy harvested from the ET). By leveraging the Lagrange duality method, we present the optimal solution to this problem in a semi-closed form. Numerical results show that the proposed wireless powered user cooperative computation design significantly improves the computation rate at the user, as compared to conventional schemes without such cooperation.


\section{System Model and Problem Formulation}\label{sec:system}

\begin{figure}
 \centering
 \includegraphics[width=3.5in]{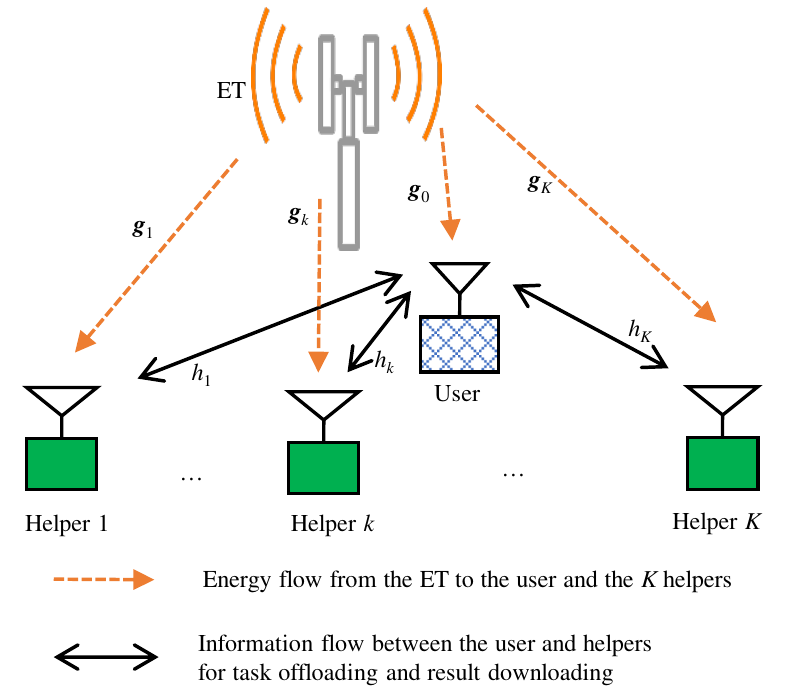}
 \caption{An illustration of the wireless powered user cooperative computation system.}\label{fig:SysMod}
\end{figure}

Consider a wireless powered cooperative computation system as shown in Fig.~\ref{fig:SysMod}, which consists of an $N$-antenna ET, a single-antenna user, and a set ${\cal K} \triangleq \{1,\ldots,K\}$ of $K$ single-antenna helpers. The ET employs the RF signal based transmit energy beamforming to simultaneously charge the user and the $K$ helpers. Relying on the harvested energy, the user can execute part of its computation tasks via local computing, and offload the remaining parts to the $K$ helpers for remote execution. In order to avoid the co-channel interference, we consider that the WPT and the multiuser communication (for task offloading and results downloading) are implemented over orthogonal frequency bands.

We focus on one particular time block with duration $T$, during which the wireless channels are assumed to remain unchanged and the user aims to maximize the computation rate (to be defined later) over this block. Furthermore, we assume that there is a central controller that can collect the global  channel state information (CSI) and the computation-related information. Therefore, the central controller can coordinate the WPT and the computation offloading.

\subsection{Energy Beamforming at ET}
First, we consider the energy beamforming from the ET to the user and helpers. Let $\bm s\in{\mathbb C}^{N\times1}$ denote the energy-bearing signal sent by the ET, and $\bm Q=\mathbb{E}[\bm s\bm s^H]\in{\mathbb C}^{N\times N}$ denote the transmit energy covariance matrix, where $\mathbb{E}[\cdot]$ denotes the expectation operation and the superscript $H$ denotes the conjugate transpose. Accordingly, the transmit power at the ET is given by $\mathbb{E}[\|\bm s\|^2]={\rm tr}(\bm Q)$, where $\|\cdot\|$ denotes the Euclidean norm of a vector and ${\rm tr}(\cdot)$ denotes the trace of a matrix. Denoting $P_{\max}$ as the maximum transmit power of the ET, it follows that ${\rm tr}(\bm Q) \leq P_{\max}$. In general, the ET can employ multiple energy beams to deliver wireless energy, i.e., $\bm Q$ can be of any rank. In particular, suppose that $r={\rm rank}(\bm Q)\leq N$, where ${\rm rank}(\bm A)$ denotes the rank of matrix $\bm A$. Accordingly, there are a total of $r$ energy beams that can be obtained by the eigenvalue decomposition (EVD) of $\bm{Q}$.

Furthermore, let $\bm g_0\in{\mathbb C}^{N\times 1}$ and $\bm g_k\in{\mathbb C}^{N\times 1}$ denote the channel vectors from the ET to the user (with index $0$ for notational convenience) and helper $k\in{\cal K}$, respectively. The received RF power at the user and helper $k$ are given by $|\bm g_0^H \bm s|^2$ and $|\bm g_k^H \bm s|^2$, respectively, where $|\cdot|$ denotes the absolute value of a scalar. As commonly adopted in the WPT literature\cite{4,5,6}, we assume a linear energy harvesting (EH) model for both the user and helpers. Consequently, the harvested energy amount by the user (with index $k=0$) or helper $k\in{\cal K}$ over this block is given by
\begin{align}
E_k = T\zeta_k\mathbb{E}[|\bm g_k^H \bm s|^2] = T \zeta_k{\rm tr}(\bm Q \bm g_k\bm g_k^H),~~k\in\{0\}\cup \mathcal K,
\end{align}
where $0 < \zeta_k \leq 1$ denotes the constant EH efficiency of the user or helper.


\subsection{User Cooperative Computing}
Next, we explain the cooperative computation between the user and helpers. Consider the partial offloading case \cite{12,13}, such that the user can arbitrarily partition the computation tasks into $(K+1)$ parts for parallel execution at the user and the helpers, respectively. We denote $\ell_{0}$ as the number of task input-bits for the user's local computing and $\ell_{k}$ as that for computation offloading from the user to helper $k\in{\cal K}$.

\begin{figure}
 \centering
 \includegraphics[width=3.5in]{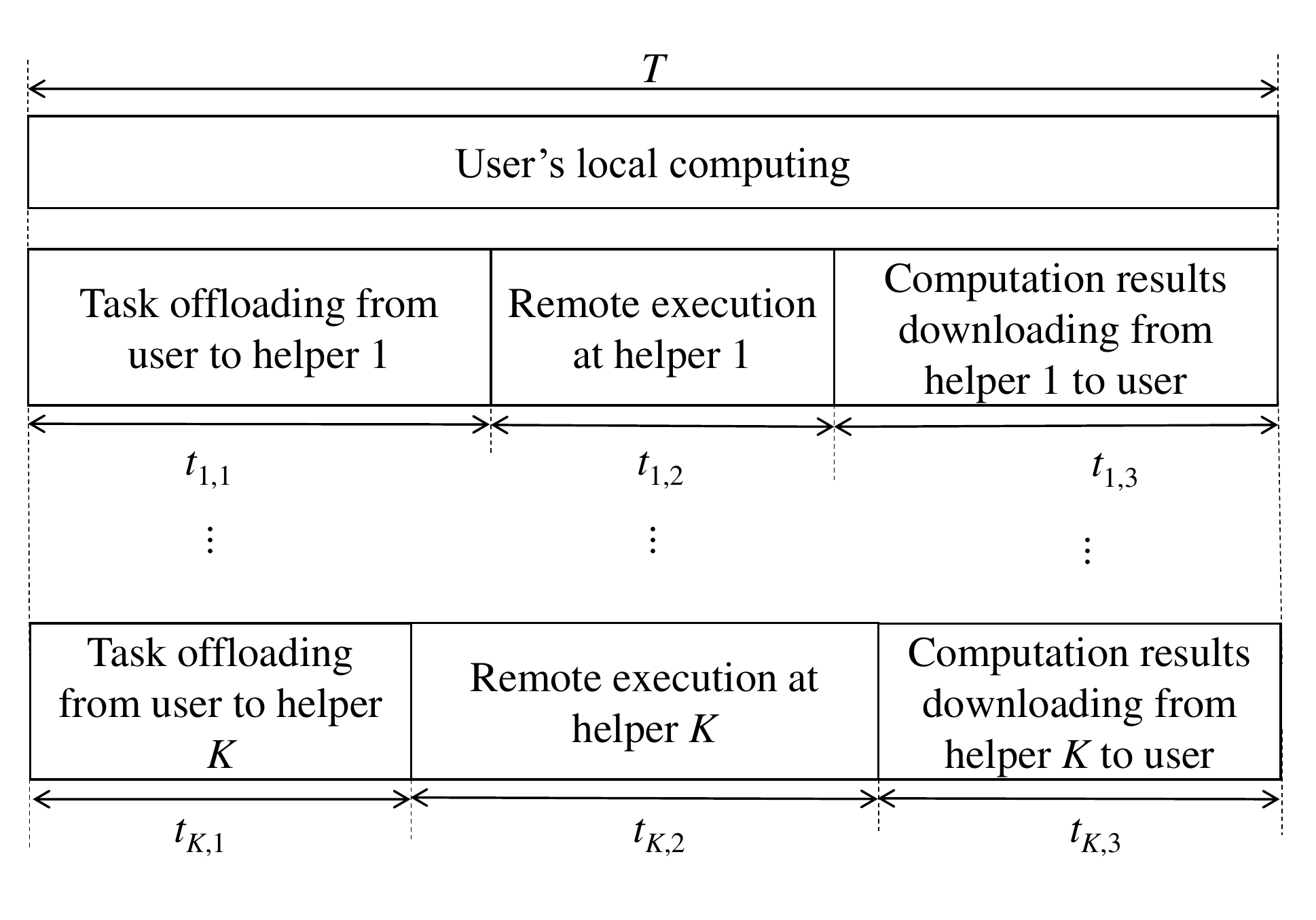}
 \caption{An illustration of the FDMA-based computation offloading protocol.}\label{fig:protocol}
\end{figure}

The computation offloading between the user and the $K$ helpers is based on the FDMA protocol, as shown in Fig.~\ref{fig:protocol}. The communication between the user and each helper is allocated with an orthogonal frequency band with bandwidth $B$. For each helper $k\in{\cal K}$, the block is divided into three time slots with durations $t_{k,1}$, $t_{k,2}$, and $t_{k,3}$, for user's task offloading to helper $k$, helper $k$'s remote computing, and computation result downloading from helper $k$ to the user, respectively. It thus follows that $\sum_{i=1}^3 t_{k,i} \leq T$.

Let $h_{k}$ denote the channel power gain between the user and helper $k\in{\cal K}$. In the first slot, the number of offloaded task input-bits from the user to helper $k$ is given as
\begin{align}
\ell_k=t_{k,1}B \log\left(1+\frac{h_kq_k}{\sigma^2_k}\right),
\end{align}
where $q_k$ denotes the transmit power of the user for offloading task to helper $k$ and $\sigma_k^2$ is the power of additive white Gaussian noise (AWGN) at helper $k\in{\cal K}$. Correspondingly, the total transmission energy consumption for the user's offloading is given by
\begin{align}
E_{0,{\rm tx}} = \sum_{k=1}^K q_kt_{k,1}=\sum_{k=1}^K t_{k,1}(2^{\frac{\ell_k}{t_{k,1}B}}-1)\frac{\sigma_k^2}{h_k}.
\end{align}

In the second slot, each helper $k\in{\cal K}$ executes $\ell_{k}$ task input-bits. Let $C_k$ denote the central process unit (CPU) cycles required for executing one input-bit of the offloaded task. To successfully execute the $C_k\ell_k$ CPU cycles, the energy consumption for helper $k$'s remote computing is given by\cite{3}
\begin{align}\label{eq.helper_loc}
E_{k,{\rm comp}} = \sum_{n=1}^{C_k\ell_k} \xi_{k} f_{k,n}^2,
\end{align}
where $\xi_k$ and $f_{k,n}$ denote the effective CPU switch capacitance and the CPU frequency for executing the $n$th CPU cycles of helper $k$, respectively. To minimize the energy consumption of helper $k$'s local computing, an identical CPU frequency should be adopted for every CPU cycle [13], i.e.,
\begin{align}\label{eq.helper_CPU}
f_{k,1}=f_{k,2}=\ldots=f_{k,C_k\ell_k}=\frac{C_k\ell_k}{t_{k,2}}.
\end{align}
Substituting \eqref{eq.helper_CPU} into \eqref{eq.helper_loc}, the energy consumption for helper $k$'s computing is re-expressed as
\begin{align}
E_{k,{\rm comp}} = \frac{\xi_kC_k^3\ell_k^3}{t^2_{k,2}},~~k\in{\cal K}.
\end{align}

In the last slot with duration $t_{k,3}$, the user downloads the corresponding computation results from each helper $k\in{\cal K}$. For ease of analysis, we assume that the size of the computation results is proportional to the size of the corresponding task input-bits, which is denoted as $\beta \ell_k$ for helper $k\in{\cal K}$, with $\beta>0$ denoting a task-specific constant\cite{3}. Let $p_k$ denote the transmit power of helper $k$ for sending the computation results to the user. We then have
\begin{align}
\beta\ell_k = t_{k,3}B \log\left(1+\frac{h_kp_{k}}{\sigma_0^2}\right),~~k\in{\cal K},
\end{align}
where $\sigma_0^2$ is the power of AWGN at the user. The corresponding transmission energy consumption for helper $k\in{\cal K}$ is given by
\begin{align}
E_{k,{\rm tx}} = p_k t_{k,3}=t_{k,3}(2^{\frac{\beta\ell_k}{t_{k,3}B}}-1)\frac{\sigma_0^2}{h_k}.
\end{align}

In addition, the user performs local computing to execute the $C_0\ell_0$ input-bits of the partitioned task over the whole duration-$T$ block, where $C_0$ is the CPU cycles required for each task input-bit at the user. Similarly, an identical CPU frequency $C_0\ell_0/T$ is adopted for each CPU cycle at the user. The resultant energy consumption for local computing at the user is given by\cite{3}
\begin{align}
E_{0,{\rm comp}}= \frac{\xi_0C_0^3\ell_0^3}{T^2},
\end{align}
where $\xi_0$ is the constant switch capacitance of the user's CPU architecture.

Furthermore, notice that the energy consumption at each of the user and the $K$ helpers is supplied by the WPT from the ET. Therefore, the user and $K$ helpers are each subject to the so-called energy neutrality constraints to achieve self-sustainable operation\cite{7}, i.e., over the particular block, the total energy consumed by each node cannot exceed the total energy harvested at that node. Therefore, we have
\begin{align}
& E_{k,{\rm tx}}+E_{k,{\rm comp}} \leq E_k
\end{align}
for all $k\in\{0\}\cup{\cal K}$.

\subsection{Problem Formulation}
In this paper, we aim to maximize the computation rate (i.e., the total number of task inputs $\sum_{k=0}^K\ell_k$ within the block) at the user for this wireless powered cooperative computation system. Towards this end, we jointly optimize the energy transmit covariance matrix $\bm Q$ at the ET, the numbers $\{\ell_k\}$ of the partitioned task input-bits, and the time allocations $\{t_{k,1},t_{k,2},t_{k,3}\}$ for the cooperative computing between the user and each helper. Mathematically, the energy-constrained computation rate maximization problem is formulated as
\begin{subequations}\label{eq.prob1}
\begin{align}
\max_{\bm Q \succeq \bm 0, \bm \ell,\bm t}~&  \sum_{k = 0}^{K} \ell_k\\
\quad \quad {\rm s.t.}~&  \sum_{k=1}^K t_{k,1}(2^{\frac{\ell_k}{t_{k,1}B}}-1)\frac{\sigma_k^2}{h_k} + \frac{\xi_0C_0^3\ell_0^3}{T^2} \notag \\
&\quad\quad\quad\quad\quad\quad\quad\quad\quad\quad   \leq T \zeta_0 {\rm tr}(\bm Q\bm g_0\bm g^H_0)\\
 & t_{k,3}(2^{\frac{\beta\ell_k}{t_{k,3}B}}-1)\frac{\sigma_0^2}{h_k}+\frac{\xi_kC_k^3\ell_k^3}{t^2_{k,2}} \notag \\
&\quad\quad\quad\quad\quad\quad  \leq T \zeta_k {\rm tr}(\bm Q\bm g_k\bm g_k^H),~\forall k\in{\cal K}\\
&\sum_{i=1}^3 t_{k,i} \leq T,~~\forall k\in{\cal K} \\
& \ell_k\geq 0,~~\forall k\in\{0\}\cup{\cal K}\\
& 0\leq t_{k,i} \leq T, ~~\forall i\in {\cal I},~k\in{\cal K}\\
& {\rm tr}(\bm Q) \leq P_{\max},
\end{align}
\end{subequations}
where $\bm \ell\triangleq [\ell_0,\ell_1,\ldots,\ell_K]^\dagger$, $\bm t\triangleq[\bm t^\dagger_1,\ldots,\bm t^\dagger_K]^\dagger$, $\bm t_k\triangleq[t_{k,1},t_{k,2},t_{k,3}]^\dagger$, and ${\cal I}\triangleq \{1,2,3\}$ are defined for notational convenience, with $[\cdot]^\dagger$ denoting the transpose of a vector. Note that problem \eqref{eq.prob1} is convex, due to the fact that the objective function is linear and all the constraints are convex. Therefore, problem \eqref{eq.prob1} can be efficiently solved by standard convex optimization techniques such as the interior-point method\cite{19}. Nevertheless, to reveal essential engineering insights, in the next section we employ the Lagrange duality method to obtain the optimal solution to problem \eqref{eq.prob1}.

\section{Optimal Solution to Problem \eqref{eq.prob1}}\label{sec:Solution}
In this section, we present an efficient algorithm for optimally solving \eqref{eq.prob1} based on the Lagrange duality method.

Let $\lambda_{0} \geq 0 $, $\lambda_k \geq 0$, $ \mu_k\geq 0$, $k \in \mathcal{K}$, and $\rho \geq 0$ denote the Lagrange multiplier associated with the constraints in (\ref{eq.prob1}b), (\ref{eq.prob1}c), (\ref{eq.prob1}d), and (\ref{eq.prob1}g), respectively. Then the partial Lagrangian of problem \eqref{eq.prob1} is expressed as
\vspace{-0.1cm}
\begin{align}
&{\cal L}(\bm Q,\bm \ell, \bm t, \bm \lambda,\bm \mu, \rho)= \notag\\
& {\rm tr}\Bigg(\Big(\sum_{k=0}^K \lambda_k T\zeta_k\bm g_k \bm g_k^H-\rho\bm I\Big)\bm Q\Bigg)+\rho P_{\max}+\sum_{k=1}^K \mu_kT\notag\\
& + \ell_0-\frac{\lambda_0\xi_0C_0^3}{T^2}\ell_0^3+\sum_{k=1}^K \Big(\ell_{k}-\frac{\lambda_0\sigma_k^2t_{k,1}}{h_k}(2^{\frac{\ell_{k}}{Bt_{k,1}}}-1)\notag\\
& -\sum_{i=1}^3 \mu_kt_{k,i}-\frac{\lambda_k\sigma_0^2t_{k,3}}{h_k}(2^{\frac{\beta \ell_{k}}{Bt_{k,3}}}-1)-\frac{\lambda_k\xi_kC_k^3\ell_k^3}{t^2_{k,2}}\Big).
\end{align}
Accordingly, the dual function is given by
\vspace{-0.1cm}
\begin{subequations}\label{eq.func_dual}
\begin{align}
{\cal G}(\bm \lambda,\bm \mu, \rho) = \max_{\bm Q\succeq \bm 0,\bm \ell, \bm t}~ &~ {\cal L}(\bm Q,\bm \ell, \bm t, \bm \lambda,\bm \mu, \rho)\\
{\rm s.t.}~ & ~(\ref{eq.prob1}\text{e})~\text{and}~(\ref{eq.prob1}\text{f}).
\end{align}
\end{subequations}
Then, the dual problem is expressed as
\vspace{-0.1cm}
\begin{subequations}\label{eq.prob_dual}
\begin{align}
\min_{\bm \lambda,\bm \mu, \rho}~&~ {\cal G}(\bm \lambda,\bm \mu, \rho)\\
{\rm s.t.}~&~ \bm \lambda > \bm 0,~~\bm \mu \geq \bm 0,~~\rho \geq 0\\
& \bm F(\bm \lambda, \rho)\preceq 0,
\end{align}
\end{subequations}
where $\bm F(\bm \lambda, \rho)$ $\triangleq$ $\sum_{k=0}^K\lambda_kT\zeta_k\bm g_k\bm g_k^H-\rho\bm{I}$ with $\bm I$ being an identity matrix of size $N\times N$. Note that the constraint of $\bm \lambda > \bm 0$ in (\ref{eq.prob_dual}b) and that in (\ref{eq.prob_dual}c) are imposed to ensure the dual function ${\cal G}(\bm \lambda, \bm \mu, \rho )$ bounded from above (as proved in Appendix A).

As problem \eqref{eq.prob1} is convex and satisfies the Slater's condition, strong duality holds between the primal problem \eqref{eq.prob1} and the dual problem \eqref{eq.prob_dual}. Therefore, we can solve problem \eqref{eq.prob1} by equivalently solving problem \eqref{eq.prob_dual}. In the following, we first evaluate the dual function ${\cal G}(\bm \lambda,\bm \mu,\rho)$ under any given  $(\bm \lambda,\bm \mu,\rho) \in {\cal X}$, where we denote $\cal X$ as the feasible set of $(\bm \lambda,\bm \mu,\rho)$ of problem \eqref{eq.prob_dual}, and then find the optimal dual variables $(\bm \lambda,\bm \mu,\rho)$ to minimize ${\cal G}(\bm \lambda,\bm\mu,\rho)$. We denote $(\bm \lambda^{\rm opt},\bm \mu^{\rm opt},\rho^{\rm opt})$ as the optimal dual solution to problem \eqref{eq.prob_dual}.

\subsection{Evaluating Dual Function ${\cal G}(\bm \lambda,\bm \mu,\rho)$}
First, we obtain the dual function ${\cal G}(\bm \lambda,\bm \mu,\rho)$ under any given  $(\bm \lambda,\bm \mu,\rho) \in {\cal X}$ by solving problem \eqref{eq.func_dual}. Problem \eqref{eq.func_dual} can be readily decomposed into the following $(K+2)$ independent subproblems, one for optimizing $\bm Q$, one for optimizing $\ell_0$, and the remaining $K$ subproblems for jointly optimizing $\ell_k$'s and $\bm t_k$'s.
\vspace{-0.2cm}
\begin{align}\label{eq.prob_sub1}
\max_{\bm Q}~{\rm tr} \Big(\bm F(\bm \lambda, \rho)\bm Q\Big)\quad {\rm s.t.}~\bm Q\succeq \bm 0
\end{align}
\vspace{-0.3cm}
\begin{align}\label{eq.prob_sub2}
\max_{\ell_0}~ \ell_0-\frac{\lambda_0\xi_0C_0^3}{T^2}\ell_0^3\quad {\rm s.t.}~ \ell_0 \geq 0
\end{align}
\begin{subequations}\label{eq.prob_sub3}
\begin{align}
\max_{\ell_k,\bm t_k} ~& ~ \ell_{k}-\frac{\lambda_0\sigma_k^2t_{k,1}}{h_k}(2^{\frac{\ell_{k}}{Bt_{k,1}}}-1)-\sum_{i=1}^3 \mu_kt_{k,i}\notag\\
& \quad\quad\quad -\frac{\lambda_k\sigma_0^2t_{k,3}}{h_k}(2^{\frac{\beta \ell_{k}}{Bt_{k,3}}}-1)-\frac{\lambda_k\xi_kC_k^3\ell_k^3}{t^2_{k,2}}\\
{\rm s.t.} ~& ~\ell_{k}\geq 0 ,~0 \leq t_{k,i}\leq T,~\forall i\in{\cal I},
\end{align}
\end{subequations}
where each subproblem $k$ in \eqref{eq.prob_sub3} is for one helper $k\in{\cal K}$.

For problem \eqref{eq.prob_sub1}, under the condition $\bm F(\bm \lambda, \rho)\preceq \bm 0$, the optimal value is zero and the optimal solution $\bm Q^{*}$ to problem \eqref{eq.prob_sub1} can be any positive semidefinite matrix in the null space of $\bm F(\bm \lambda, \rho)$. Here, we set $\bm Q^{*} = \bm 0$ for the purpose of evaluating the dual function ${\cal G}(\bm \lambda,\bm \mu, \rho)$.

As for the optimal $\ell_0^*$ of problem \eqref{eq.prob_sub2}, under given $\lambda_0$, the objective function in \eqref{eq.prob_sub2} is concave with respect to $\ell_0$. Therefore, based on the first-order derivative condition\cite{19}, we have
\begin{align}\label{eq.opt_l0}
\ell_0^* = \frac{T}{\sqrt{\lambda_0\xi_0C_0^3}}.
\end{align}
For the $k$th problem in \eqref{eq.prob_sub3}, it is convex and satisfies the Slater's condition. Based on the Karush-Kuhn-Tucker (KKT) conditions\cite{19}, we obtain the optimal solution of $\ell_k$ and $\bm t_k$ to problem \eqref{eq.prob_sub3} in a semi-closed form, as stated in the following lemma.
\begin{lemma}\label{lem.t_l}
The optimal solution of $\ell^*_{k}$ and $\mv t^*_k$ to problem \eqref{eq.prob_sub3} is given by
\begin{align*}
\ell^*_{k}
\begin{cases}
=0,~&{\rm if}~ {\cal M}(\bm r_k^*,\lambda_0,\lambda_k,\mu_k)< 0\\
\in[0,\min_{i\in\{1,2,3\}}r^*_{k,i}T],~&{\rm if}~ {\cal M}(\bm r_k^*,\lambda_0,\lambda_k,\mu_k)= 0\\
=\min_{i\in\{1,2,3\}}r^*_{k,i}T, ~&{\rm if}~ {\cal M}(\bm r_k^*,\lambda_0,\lambda_k,\mu_k)> 0
\end{cases}
\end{align*}
and $t^*_{k,i}=\ell^*_k/r^*_{k,i}$ for all $i\in{\cal I}$, respectively, where $\bm r^*_k\triangleq [r_{k,1}^*,r_{k,2}^*,r_{k,3}^*]^\dagger$ with
\begin{subequations}\label{eq.opt_rate}
\begin{align}
r_{k,1}^* &=  \frac{B}{\ln2} \Big( 1+{\cal W}\Big(\frac{\mu_kh_k}{\lambda_0\sigma_k^2e}-\frac{1}{e}\Big)\Big)\\
r_{k,2}^* & = \frac{1}{C_k}\Big(\frac{\mu_k}{2\lambda_k\xi_k}\Big)^{\frac{1}{3}}\\
r_{k,3}^* &= \frac{B}{\beta\ln2}\Big(1+ {\cal W}\Big(\frac{\mu_kh_k}{\lambda_k\sigma_0^2e}-\frac{1}{e}\Big)\Big)
\end{align}
\end{subequations}
and ${\cal M}(\bm r_k^*,\lambda_0,\lambda_k,\mu_k) \triangleq$
\begin{align}
&1-\frac{\lambda_0\sigma_k^2}{h_kr^*_{k,1}}(2^{\frac{r^*_{k,1}}{B}}-1)-\sum_{i=1}^3 \frac{\mu_k}{r^*_{k,i}} \notag \\
&\quad\quad\quad\quad -\frac{\lambda_k\sigma_0^2}{h_k r^*_{k,3}}(2^{\frac{\beta r^*_{k,3}}{B}}-1) -\lambda_k\xi_kC_k^3r_{k,2}^{*2}
\end{align}
for any $k\in{\cal K}$. Note $e$ is termed Euler's number and that ${\cal W}(\cdot)$ in \eqref{eq.opt_rate} is the Lambert ${\cal W}$ function\cite{18}.
\end{lemma}
\begin{IEEEproof}
See Appendix B.
\end{IEEEproof}

As stated in Lemma~\ref{lem.t_l}, if $G(\bm r^*_k,\lambda_0,\lambda_k, \mu_k)=0$, then $\ell_{k,i}^*\in[0,T]$ is generally not a unique solution to \eqref{eq.prob_sub3}. In this case, we set $\ell_{k}^*=0$, $k\in{\cal K}$, to facilitate the dual function evaluation. An additional procedure will be employed in Section III-C to retrieve the optimal primal $\ell_k^{\rm opt}$'s, together with $t_{k,i}^{\rm opt}=\ell_{k}^{\rm opt}/r^{\rm opt}_{k,i}$ for all $i\in{\cal I}$ and $k\in{\cal K}$.



By combining the optimal $\bm Q^*=\bm 0$, $\ell_0^*$ in \eqref{eq.opt_l0}, and Lemma \ref{lem.t_l}, the dual function ${\cal G}(\bm \lambda,\bm \mu,\rho)$ can be readily evaluated under given $(\bm \lambda,\bm \mu,\rho)\in{\cal X}$.

\subsection{Obtaining Optimal $(\mv \lambda^{\rm{opt}},\mv \mu^{\rm{opt}},\rho^{\rm{opt}})$ to Minimize ${\cal G}(\bm \lambda,\bm \mu, \rho)$}
Generally, the dual function ${\cal G}(\bm \lambda,\bm \mu, \rho)$ in \eqref{eq.func_dual} is convex but non-differentiable. As a result, the optimal dual solution $(\bm \lambda^{\rm opt},\bm \mu^{\rm opt}, \rho^{\rm opt})$ to problem \eqref{eq.prob_dual} can be obtained by subgradient based methods such as the ellipsoid method [19]. To begin with, we choose a given  $(\mv \lambda,\mv \mu,\rho) \in {\cal X}$ as the center of the initial ellipsoid and set its volume to be sufficiently large to contain the optimal $(\bm \lambda^{\rm opt},\bm \mu^{\rm opt},\rho^{\rm {opt}})$. Then, at each iteration, we update the dual variables $(\bm \lambda,\bm \mu,\rho)$ based on the subgradients of both the objective function and the constraints in problem \eqref{eq.prob_dual}, and accordingly establish a new ellipsoid with reduced volume. When the ellipsoid volume is below a certain threshold, the iteration terminates and the ellipsoid center is chosen to be the optimal $(\bm \lambda^{\rm opt},\bm \mu^{\rm opt},\rho^{\rm {opt}})$.

To implement the ellipsoid method, it remains to determine the subgradients of both the objective function and constraints. For the objective function in (\ref{eq.prob_dual}a), the subgradient with respect to $(\bm \lambda,\bm \mu,\rho)$ is given as
\vspace{-0.2cm}
\begin{align*}\label{eq.subg}
\Big[ &T \zeta_0 {\rm tr}(\bm Q\bm g_0\bm g^H_0)- \sum_{k=1}^K t_{k,1}(2^{\frac{\ell_k}{t_{k,1}B}}-1)\frac{\sigma_k^2}{h_k} - \frac{\xi_0C_0^3\ell_0^3}{T^2},\notag \\
&T \zeta_1 {\rm tr}(\bm Q\bm g_1\bm g_1^H)-\frac{\sigma_0^2t_{1,3}}{h_1}(2^{\frac{\beta\ell_1}{t_{1,3}B}}-1)- \frac{\xi_1C_1^3\ell_1^3}{t^2_{1,2}}, \ldots,\notag\\
&T \zeta_K {\rm tr}(\bm Q\bm g_K\bm g_K^H)-\frac{\sigma_0^2t_{K,3}}{h_k}(2^{\frac{\beta\ell_K}{t_{K,3}B}}-1)- \frac{\xi_KC_K^3\ell_K^3}{t^2_{K,2}}, \notag\\
&T-\sum_{i=1}^3 t_{1,i},\ldots,T-\sum_{i=1}^3 t_{K,i}, P_{\max}-{\rm tr}(\bm Q)\Big]^\dagger.
\end{align*}
The subgradients for the constraints in (\ref{eq.prob_dual}b) are given by $\bm e_{k}$ for all $k\in \{0\}\cup{\cal K}$ and $\bm e_{K+k}$ for all $k\in{\cal K}$, respectively, where $\bm e_j\in {\mathbb R}^{(2K+2)\times1}$ is the standard unit vector with one in the $j$th entry and zeros elsewhere. By using this together with \eqref{eq.subg}, the ellipsoid method can be applied to efficiently update $(\bm \lambda,\bm \mu,\rho)$ towards $(\bm \lambda^{\rm opt},\bm \mu^{\rm opt}, \rho^{\rm {opt}})$ for problem \eqref{eq.prob_dual}.

\subsection{Finding Optimal Primal $(\bm Q^{\rm opt},\bm \ell^{\rm opt},\bm t^{\rm opt})$}
With the optimal dual solution $(\bm \lambda^{\rm opt},\bm \mu^{\rm opt}, \rho^{\rm opt})$, it remains to determine the optimal primal solution to problem \eqref{eq.prob1}. Specifically, substituting $\bm \lambda^*$ and $\bm \mu^*$ with $\bm \lambda^{\rm opt}$ and $\bm \mu^{\rm opt}$, respectively, we obtain the optimal $\ell_{0}^{\rm opt}$ and $\bm r_{k}^{\rm opt}$. Due to the non-uniqueness of $\ell_{k}^*$'s and $\bm Q^{*}$, one cannot obtain $\ell_{k}^{\rm opt}$'s and $\bm Q^{\rm opt}$ directly here but resort to an additional procedure. By substituting $\ell_0^{\rm opt}$ and $r_{k,i}^{\rm opt}=\ell_k/t_{k,i}$, $i\in{\cal I}$, $k\in{
\cal K}$, in problem \eqref{eq.prob1}, we solve the following semidefinite program (SDP) problem to obtain the optimal primal $\bm Q^{\rm opt}$ and $\ell_k^{\rm{opt}}$'s:
\begin{subequations}\label{prob.final}
\begin{align}
\max_{\substack{\bm Q \succeq \bm 0 \\ \{\ell_k,k\in{\cal K}\}}} &~ \sum_{k = 1}^{K} \ell_k \\
{\rm s.t.}~ & \sum_{k=1}^K (2^{\frac{r_{k,1}^{\rm opt}}{B}}-1)\frac{\sigma_k^2\ell_{k}}{h_kr_{k,1}} + {\xi_0^{\frac{1}{2}}C_0^{\frac{3}{2}}T}/{\sqrt{\lambda_0^{\rm opt}}} \notag\\
&\quad\quad\quad\quad\quad \leq T \zeta_0 {\rm tr}(\bm Q\bm g_0\bm g^H_0),~\forall k\in{\cal K}\\
& (2^{\frac{\beta r_{k,3}^{\rm opt}}{B}}-1)\frac{\ell_{k}\sigma_0^2}{h_kr_{k,3}}+\xi_kC_k^3\ell_{k}r_{k,2}^{\rm opt}\notag\\
&\quad\quad\quad\quad\quad \leq T \zeta_k {\rm tr}(\bm Q\bm g_k\bm g_k^H),~\forall k\in{\cal K} \\
&\sum_{k = 1}^{K}\frac{\ell_{k}}{r_{k,i}^{\rm opt}} \leq T,~\ell_k\geq 0,~\forall k\in{\cal K} \\
&{\rm tr}(\bm Q) \leq P_{\max}.
\end{align}
\end{subequations}
Note that the SDP in \eqref{prob.final} can be efficiently solved via CVX toolbox [19]. With $\ell_k^{\rm opt}$'s obtained, we have the optimal primal $t_{k,i}^{\rm opt}=\ell_k^{\rm opt}/r^{\rm opt}_{k,i}$, $i\in{\cal I}$, $k\in{\cal K}$, to problem \eqref{eq.prob1}. Then, by combining $\bm Q^{\rm opt}$, $\ell_k^{\rm opt}$'s, and $\bm t^{\rm opt}$ here, together with $\ell_{0}^{\rm opt}$, the optimal solution to problem \eqref{eq.prob1} is finally found. 


\section{Numerical Results}\label{results}

In this section, we present numerical results to evaluate the performance of the proposed wireless powered user cooperative computing design, as compared with the following two benchmark schemes.

\subsubsection{Local computing only} The user accomplishes its computation task by local computing only. This scheme corresponds to solving problem \eqref{eq.prob1} by setting $\ell_k = 0$ and $\ t_{k} = 0$ for all $k\in{\cal K}$.

\subsubsection{Equal time allocation for offloading} The time durations for the task offloading from the user to each helper, the task execution at each helper, and the computation result downloading from different helpers to the user are equally allocated. This scheme corresponds to solving problem \eqref{eq.prob1} by setting $t_{k,1}= t_{k,2} = t_{k,3} = T/3$ for all $k\in{\cal K}$.

In this simulation, we set the number of antennas at the ET as $N= 4$ and the number of helpers as $K=3$. For both the user and the helpers, we set the EH efficiency as $\zeta_k = 0.6$, the switch capacitance as $\xi_k = 10^{-28}$, the required CPU cycles per bit as $C_k=10^3$\cite{13}, the receive noise power $\sigma^2_k=10^{-9}$ Watt (W), $k \in \{0\}\cup\mathcal{K}$. The bandwidth used for the communication between the user and each helper is set as $B = 1$ MHz. All channels are modeled as independent Rayleigh fading with an average power gain of ${\rm PL}_{0}\times d_{k}^{-3}$, where ${\rm PL}_{0}= 10^{-3}$ is the channel power gain at a reference distance of 1 meter (m), $d_{k}$ denotes the distance from the user to helper $k\in{\cal K}$, and the path-loss exponent is assumed to be 3.

\begin{figure}
    \centering
	\includegraphics[width = 3.8in]{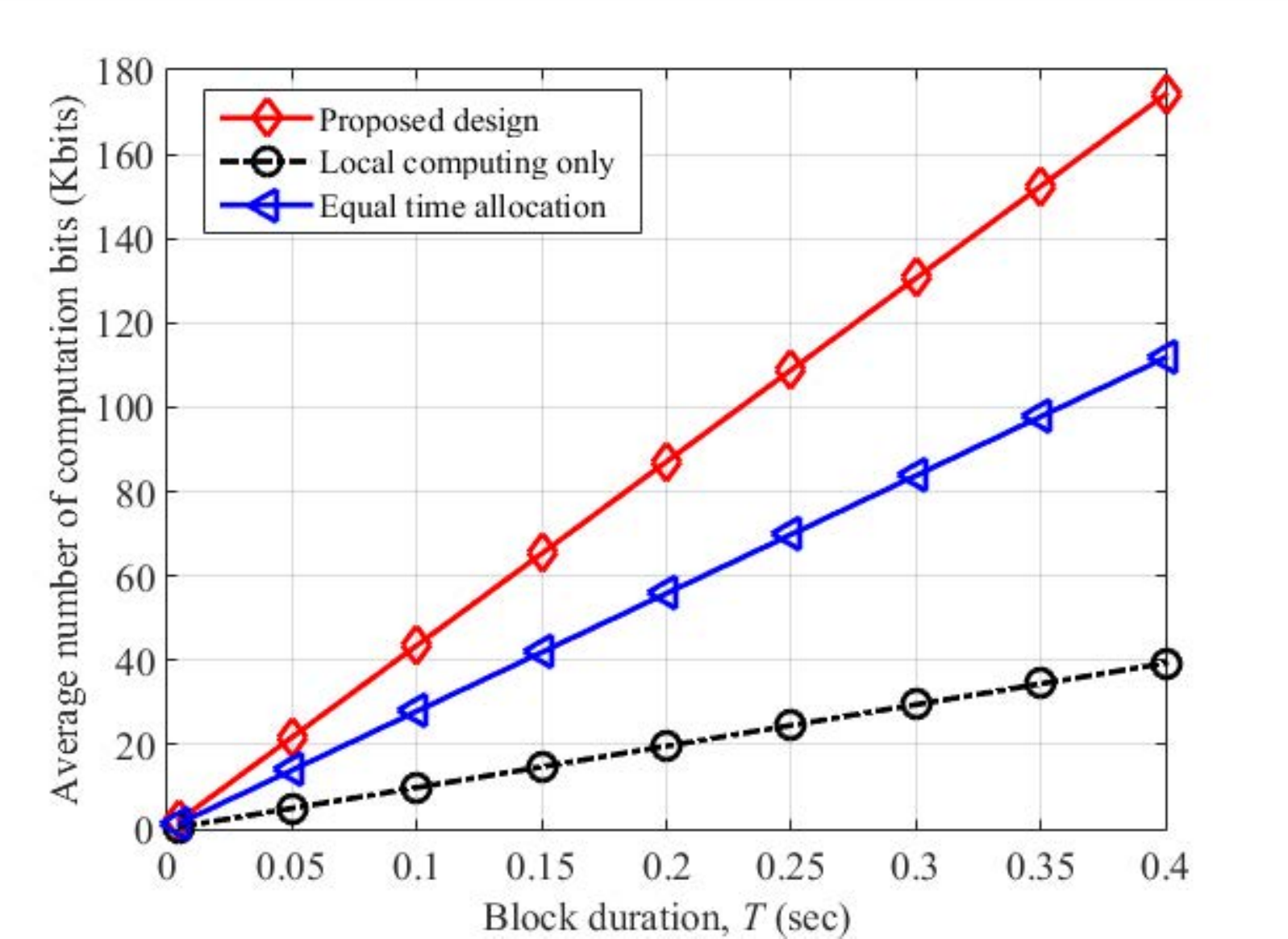}
    \caption{The average number of computation bits of the user versus the block duration $T$. }\label{fig:vs_T}
\end{figure}

\begin{figure}
 \centering
\includegraphics[width=3.8in]{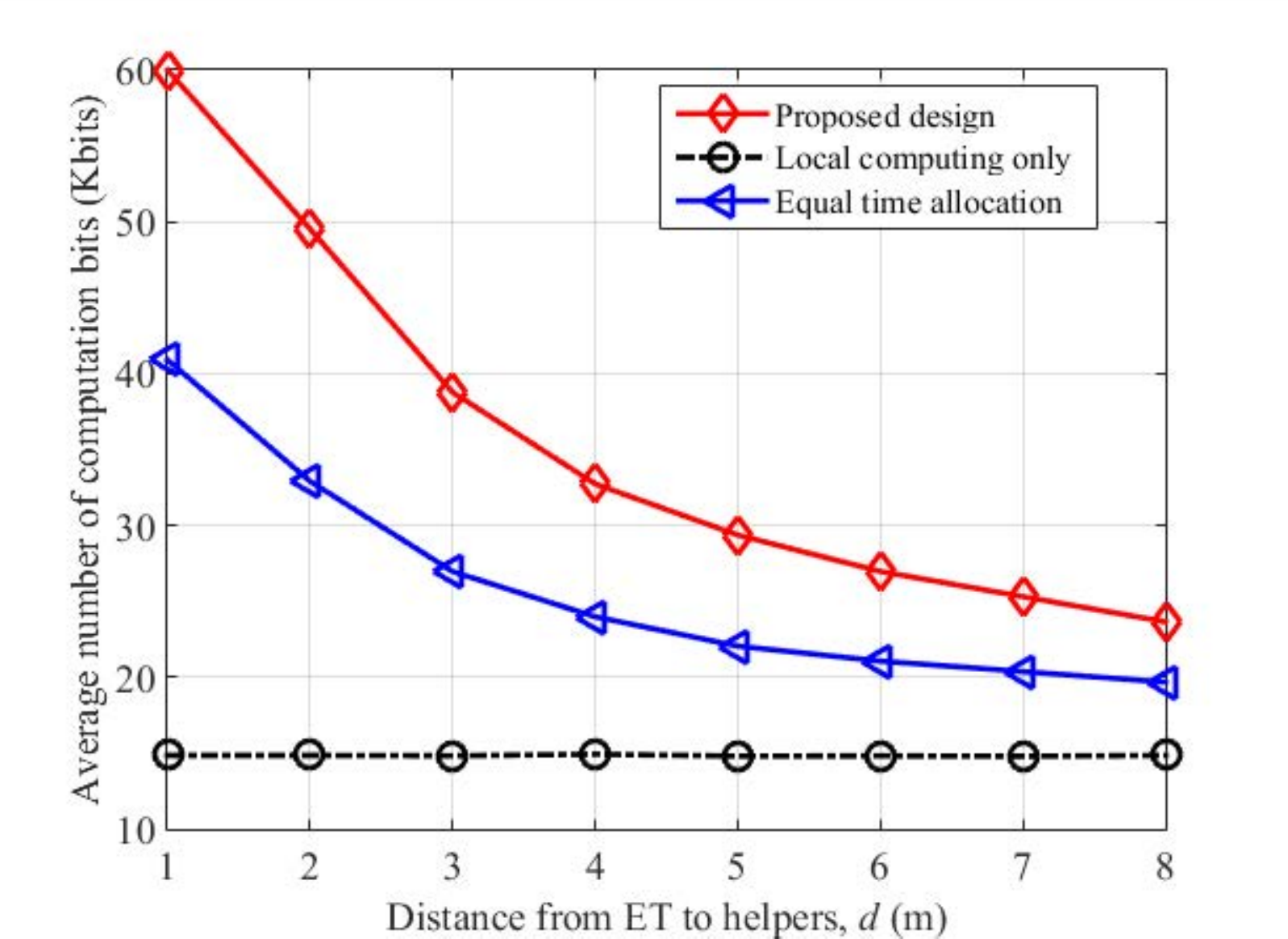}
   \caption{The average number of computation bits of the user versus the distance from the ET to the helpers.}	
   \label{fig:vs_g}
\end{figure}

\begin{figure}
   \centering
\includegraphics[width=3.8in]{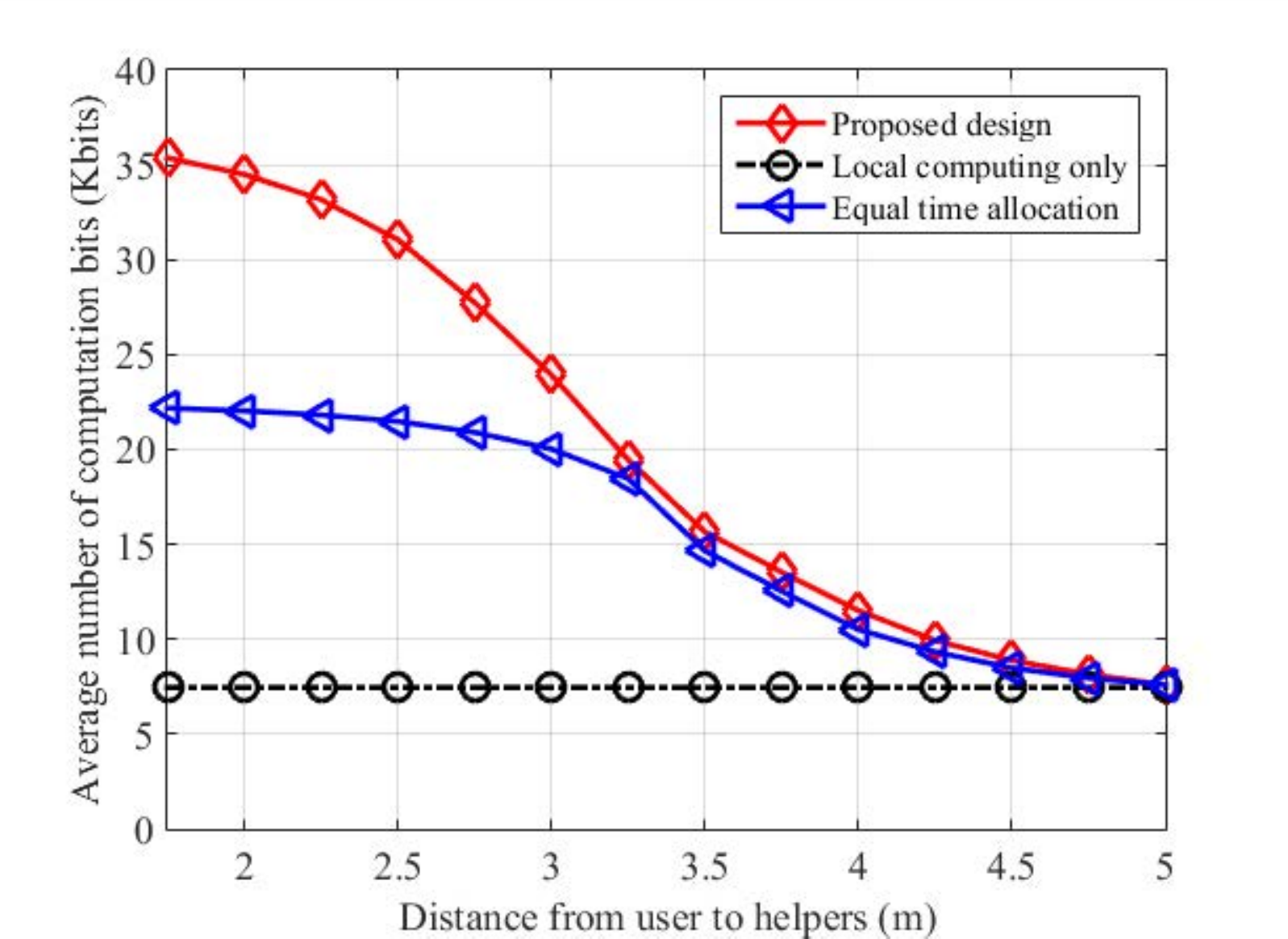}
   \caption{The average number of computation bits of the user versus the distance from the user to the helpers.}	
   \label{fig:vs_h}
\end{figure}

Fig. \ref{fig:vs_T} shows the average number of computation bits of the user versus the block duration $T$, where the distances from the user and the helpers are set to be $d_1=2$ m, $d_2=3$ m, and $d_3=5$ m, respectively. It is observed that the proposed design outperforms the two benchmark schemes. This shows the merit of the joint resource allocation in our design for performance optimization.

Fig. \ref{fig:vs_g} shows the average number of computation bits of the user versus the distance $d$ between the ET and the helpers, where the locations for the ET and the user are fixed and the block duration is $T=0.1$ sec. The proposed design is observed to outperform all the benchmark schemes. As the distance $d$ increases, the average numbers of computation bits achieved by the proposed design and the equal-time-allocation scheme both reduce significantly. This is due to the fact that the harvested energy at the helpers becomes smaller, and thus the user cooperative computation gain decreases.

Fig. \ref{fig:vs_h} shows the average number of computation bits of the user versus the distance between the user and the helpers, where the locations of the ET and the user are fixed and the block duration is $T=10^{-3}$ sec. It is observed that the performance gain of the proposed design reduces as the distance between the user and helpers increases.

\section{Conclusion}\label{conclusion}
In this paper, we investigated a novel wireless powered user cooperative computation design for MEC systems, where nearby wireless devices are exploited as helpers that can opportunistically harvest wireless energy for cooperatively computing active users' tasks. Specifically, we developed an efficient design framework to maximize the computation rate at the user  within a given block subject to the energy neutrality constraints at the user and helpers. Based on the Lagrange duality method, we obtained the optimal semi-closed solution to this problem. Numerical results showed the substantial performance gain of the proposed wireless powered user cooperative computing design, over the benchmark schemes without such cooperation. It is our hope that the proposed design can provide a new viable means to efficiently improve the computation performance of wireless devices in future IoT networks that integrate wireless communication, computation, and power in a unified manner.

\section*{Appendices}
\subsection{Proof of Conditions $\bm F(\bm \lambda, \rho)\preceq \bm 0$ and $\bm\lambda > \bm 0$}
The condition $\bm F(\bm \lambda, \rho)\preceq \bm 0$ can be verified by contradiction. Assume that $\bm F(\bm \lambda, \rho)$ is not negative semidefinite. Denote by $\bm \nu \in {\mathbb C}^{N\times 1}$ an eigenvector corresponding to one positive eigenvalue of $\bm F(\bm \lambda, \rho)$. By setting $\bm Q$ = $\tau\bm\nu\bm\nu^{H}$ $\geq$ 0 with $\tau$ going to positive infinity, it follows that
\begin{align}
&\lim_{\tau \rightarrow +\infty} { \rm tr}\Big(\bm F(\bm \lambda, \rho)\bm Q\Big) = \lim_{\tau \rightarrow +\infty} \tau\bm\nu^{H}\bm F(\bm \lambda, \rho)\bm\nu = +\infty,
\end{align}
which in turn implies that the value ${\cal G}(\bm \lambda,\bm \mu, \rho)$ in \eqref{eq.prob_dual} is unbounded from above over $\bm Q$ $\succeq$ 0. Hence, to ensure that ${\cal G}(\bm \lambda,\bm \mu, \rho)$ is bounded, it requires that $\bm F(\bm \lambda, \rho)\preceq 0$.

Under the condition of $\lambda_k=0$, $k\in\{0\}\cup{\cal K}$, since that $\ell_k$ is the dominant term in the expression ${\cal L}(\bm Q,\bm \ell,\bm t,\bm \lambda,\bm \mu, \rho)$, the value of ${\cal G}(\bm \lambda,\bm \mu, \rho)$ becomes positive infinity as $\ell_{k}$ approaches positive infinity. Thus, it follows that $\bm \lambda > \bm 0$.

\subsection{Proof of Lemma \ref{lem.t_l} }\label{appendix:1}
Given $(\bm \lambda,\bm \mu,\rho)\in{\cal X}$, we solve problem \eqref{eq.prob_sub3} for any $k\in{\cal K}$. First, define $\bar{\bm \theta}_k \triangleq [\bar{\theta}_{k,1},\bar{\theta}_{k,2},\bar{\theta}_{k,3}]^\dagger$ and $\underline{\bm \theta}_k \triangleq [\underline{\theta}_{k,1},\underline{\theta}_{k,2},\underline{\theta}_{k,3}]^\dagger$. The Lagrangian of problem \eqref{eq.prob_sub3} is given by ${\cal L}_k(\ell_k,\bm t_k,\gamma_k,\bar{\bm \theta}_{k},\underline{\bm \theta}_k)\triangleq$
\begin{align}
&\ell_k-\frac{\lambda_0\sigma_k^2t_{k,1}}{h_k}(2^{\frac{\ell_k}{t_{k,1}B}}-1)
-\frac{\lambda_k\sigma_0^2t_{k,3}}{h_k}(2^{\frac{\beta\ell_k}{t_{k,3}B}}-1)+\gamma_k\ell_k \notag \\
& -\lambda_k\frac{\xi_kC_k^3\ell_k^3}{t_{k,2}^2} +\sum_{i=1}^3 \Big(\bar{\theta}_{k,i}(T-t_{k,i})+\underline{\theta}t_{k,i}-\mu_k t_{k,i}\Big),
\end{align}
where $\gamma_k$, $\bar{\theta}_{k,i}$, and $\underline{\theta}_{k,i}$, $i\in{\cal I}$, are the non-negative Lagrange multipliers associated with $\ell_k\geq 0$, $t_{k,i}\leq T$, and $t_{k,i}\geq 0$, respectively. Based on the KKT conditions, the necessary and sufficient conditions for the optimal primal-dual point $(\ell^*_k, \bm t_k^*,\gamma_k^*, \bar{\bm \theta}^*_k, \underline{\bm \theta}^*_k)$ are\cite{19}
\begin{subequations}\label{eq.sub_kkt}
\begin{align}
&\ell^*_k\geq 0,~0 \leq t^*_{k,i} \leq T, ~~\forall i\in{\cal I}\\
&\gamma^*_k\geq 0,~\bar{\theta}^*_{k,i}\geq 0,~\underline{\theta}^*_{k,i} \geq 0,~~\forall i\in{\cal I}\\
&\gamma^*_k\ell^*_k = 0,~\bar{\theta}^*_{k,i}(T-t^*_{k,i})=0,~\underline{\theta}^*_{k,i}t^*_{k,i}=0,~~\forall i\in{\cal I}\\
&1-\frac{\lambda_0\sigma_k^2\ln 2}{Bh_k}2^{\frac{r_{k,1}^*}{B}}-\frac{\beta \lambda_k\sigma_0^2\ln 2}{B h_k}2^{\frac{\beta r^*_{k,3}}{B}}-{3\lambda_k\xi_kC_k^3 r_{k,i}^{*2}} \notag \\
&\quad\quad\quad\quad\quad\quad\quad\quad\quad\quad\quad\quad\quad\quad\quad\quad\quad+\gamma^*_k = 0 \\
& \frac{\lambda_0\sigma_k^2}{h_k}2^{\frac{r_{k,1}^*}{B}}\Big( \frac{r_{k,1}^*}{B}\ln2 -1\Big)+\frac{\lambda_0\sigma_k^2}{h_k}-\bar{\theta}^*_{k,1}+\underline{\theta}^*_{k,1}-\mu_k = 0\\
& 2\lambda_k\xi_kC_k^3 r^{*3}_{k,3} -\bar{\theta}^*_{k,2}+\underline{\theta}^*_{k,2} -\mu_k= 0\\
& \frac{\lambda_k\sigma_0^2}{h_k}2^{\frac{\beta r_{k,3}^*}{B}}\Big(\frac{\beta r_{k,3}^*}{B}\ln2-1\Big)+\frac{\lambda_k\sigma_0^2}{h_k}-\bar{\theta}^*_{k,3}+\underline{\theta}^*_{k,3}\notag\\
&\quad\quad\quad\quad\quad\quad\quad\quad\quad\quad\quad\quad\quad\quad\quad\quad\quad-\mu_k=0,
\end{align}
\end{subequations}
where $r^*_{k,i} \triangleq \ell_i^*/t_{k,i}^*$ for all $i\in{\cal I}$. The left-hand-side (LHS) terms of (\ref{eq.sub_kkt}d)--(\ref{eq.sub_kkt}g) are the first-order derivatives of ${\cal L}_k$ with respect to $\ell^*_k$, $t^*_{k,1}$, $t^*_{k,2}$, and $t^*_{k,3}$, respectively. From (\ref{eq.sub_kkt}b) and (\ref{eq.sub_kkt}e), it follows that
\begin{align}\label{eq.kkt_r1}
(\frac{r_{k,1}^*}{B}\ln2-1)e^{(\frac{r_{k,1}^*}{B}\ln2-1)} = \frac{\mu_kh_k}{\lambda_0\sigma_k^{2} e}-\frac{1}{e}.
\end{align}
For the function $y=xe^x$ of $x>0$, its inverse function can be shown to be $x={\cal W}(y)$\cite{18}. Therefore, based on \eqref{eq.kkt_r1} and some simple manipulation, we have
\begin{align}
r_{k,1}^* = \frac{B}{\ln2}\Big(1+{\cal W}\Big(\frac{\mu_k^*h_k}{\lambda_0\sigma_k^2e}-\frac{1}{e}\Big)\Big).
\end{align}
Based on (\ref{eq.sub_kkt}b) and (\ref{eq.sub_kkt}f), it follows that
\begin{align}
r_{k,2}^* = \frac{1}{C_k}\big(\frac{\mu_k}{2\lambda_k\xi_k}\big)^{\frac{1}{3}}.
\end{align}
Similarly, from (\ref{eq.sub_kkt}b) and (\ref{eq.sub_kkt}g), we have
\begin{align}
r_{k,3}^* = \frac{B}{\beta\ln2}\Big(1+{\cal W}\Big(\frac{\mu_k^*h_k}{\lambda_k\sigma_0^2e}-\frac{1}{e}\Big)\Big).
\end{align}
To determine the optimal $\ell^*_k$ to problem \eqref{eq.prob_sub3}, we substitute $\bm r_{k}^*\triangleq [r_{k,1}^*,r_{k,2}^*,r_{k,3}^*]^\dagger$ into problem \eqref{eq.prob_sub3} and then obtain the following equivalent linear program (LP):
\begin{subequations}\label{prob.ell_k}
\begin{align}
\max_{\ell_i}~&\ell_k {\cal M}(\bm r_k^*,\lambda_0,\lambda_k,\mu_k)\\
{\rm s.t.}~ &~0 \leq \ell_k \leq r^*_{k,i}T,~\forall i\in{\cal I},
\end{align}
\end{subequations}
where ${\cal M}(\bm r_k^*,\lambda_0,\lambda_k,\mu_k) \triangleq 1-\frac{\lambda_0\sigma_k^2}{h_kr^*_{k,1}}(2^{\frac{r^*_{k,1}}{B}}-1)-\sum_{i=1}^3 \frac{\mu_k}{r^*_{k,i}}-\frac{\lambda_k\sigma_0^2}{h_k r^*_{k,3}}(2^{\frac{\beta r^*_{k,3}}{B}}-1)-\lambda_k\xi_kC_k^3r_{k,2}^{*2}$ for any $k\in{\cal K}$. From the solution to the LP \eqref{prob.ell_k}, it follows that the optimal $\ell_k^*$ to problem \eqref{eq.prob_sub3} is given by
\begin{align*}
\ell_k^*
\begin{cases}
=0,~&~{\rm if}~{\cal M}(\bm r_k^*,\lambda_0,\lambda_k,\mu_k)<0\\
\in [0,\min_{i\in\{1,2,3\}} r^*_{k,i}T],~&~{\rm if}~{\cal M}(\bm r_k^*,\lambda_0,\lambda_k,\mu_k)=0\\
=\min_{i\in\{1,2,3\}}r^*_{k,i}T, ~&~{\rm if}~ {\cal M}(\bm r_k^*,\lambda_0,\lambda_k,\mu_k)> 0.
\end{cases}
\end{align*}
Next, the optimal $t^*_{k,i}$ of problem \eqref{prob.ell_k} is readily obtained as $t^*_{k,i}=\ell^*_{k}/r^*_{k,i}$ for all $i\in{\cal I}$.

\ifCLASSOPTIONcaptionsoff
  \newpage
\fi







\begin{thebibliography}{1}

\bibliographystyle{IEEEbib}
\bibitem{1}
S. Barbarossa, S. Sardellitti, and P. Di Lorenzo, ``Communicating while computing: Distributed mobile cloud computing over 5G heterogeneous networks,'' {\it IEEE Signal Process. Mag.}, vol. 31, pp. 45--55, Nov. 2014. 

\bibitem{2}
Y. Hu, M. Patel, D. Sabella, N. Sprecher, and V. Young, ``Mobile edge computing: A key technology towards 5G,'' ETSI, Sophia Antipolis, France, White Paper 11, 2015. [Online]. Available: \url{http://www.etsi.org/images/files/ETSIWhitePapers/etsi/5g.pdf}

\bibitem{3}	
Y. Mao, C. You, J. Zhang, K. Huang, and K. B. Letaief, ``A survey on mobile edge computing: The communication perspective,'' {\it IEEE Commun. Survey Tuts.}, vol. 19, no. 4, pp. 2322--2358, 4th Quart. 2017.

\bibitem{4}
J. Xu and R. Zhang, ``Energy beamforming with one-bit feedback,'' {\it IEEE Trans. Signal Process.}, vol. 62, no. 20, pp. 5370--5381, Oct. 2014.

\bibitem{5}
J. Xu, Y. Zeng, and R. Zhang, ``UAV-enabled wireless power transfer: Trajectory design and energy optimization,'' {\em IEEE Trans. Wireless Commun.}, vol. 17, no. 8, pp. 5092--5106, Aug. 2018.

\bibitem{6}
Y. Huang and B. Clerckx, ``Waveform design for wireless power transfer with limited feedback,'' {\it IEEE Trans. Wireless Commun.}, vol. 17, no. 1, pp. 415--429, Jan. 2018

\bibitem{7}
Y. Zeng, B. Clerckx, and R. Zhang, ``Communications and signals design for wireless power transmission,'' {\it IEEE Trans. Commun.}, vol. 65, no. 5, pp. 2264--2290, May 2017.

\bibitem{8}
J. Xu, L. Liu, and R. Zhang, ``Multiuser MISO beamforming for simultaneous wireless information and power transfer,'' {\it IEEE Trans. Signal Process.}, vol. 62, no. 3, pp. 4798--4810, Sep. 2014.

\bibitem{9}
S. Bi, C. K. Ho, and R. Zhang, ``Wireless powered communication: Opportunities and challenges,'' {\it IEEE Commun. Mag.}, vol. 53, no. 4, pp. 117--125, Apr. 2015.

\bibitem{10}
X. Lu, P. Wang, D. Niyato, D. I. Kim, and Z. Han, ``Wireless networks with RF energy harvesting: A contemporary survey,'' {\it IEEE Commun. Surveys Tuts.}, vol. 17, no. 2, pp. 757--789, 2nd Quart. 2015.

\bibitem{11}
D. W. K. Ng, E. S. Lo, and R. Schober, ''Wireless information and power transfer: Energy efficiency optimization in OFDMA systems,'' {\it IEEE Trans. Wireless Commun.}, vol. 12, pp. 6352--6370, Dec. 2013. 

\bibitem{12}
C. You, K. Huang, and H. Chae, ``Energy efficient mobile cloud computing powered by wireless energy transfer,'' {\it IEEE J. Sel. Areas Commun.}, vol. 34, no. 5, pp. 1757--1771, May 2016.

\bibitem{13}
F. Wang, J. Xu, X. Wang, and S. Cui, ``Joint offloading and computing optimization in wireless powered mobile-edge computing systems,'' {\it IEEE Trans. Wireless Commun.}, vol. 17, no. 3, pp. 1784--1797, Mar. 2018.

\bibitem{APCC17}
F. Wang, ``Computation rate maximization for wireless powered mobile edge computing,'' in {\em Proc. APCC}, Perth, Australia, Dec. 2017, pp. 1--6.

\bibitem{14}
S. Bi and Y. J. A. Zhang, ``Computation rate maximization for wireless powered mobile-edge computing with binary computation offloading,'' {\it IEEE Trans. Wireless Commun.}, vol. 17, no. 6, pp. 4177--4190, Jun. 2018.

\bibitem{15}
X. Hu, K.-K. Wong, and K. Yang, ``Wireless powered cooperation-assisted mobile edge computing,'' {\it IEEE Trans. Wireless Commun.}, vol. 17, no. 4, pp. 2375--2388, Apr. 2018.

\bibitem{16}
X. Cao, F. Wang, J. Xu, R. Zhang, and S. Cui, ``Joint computation and communication cooperation for mobile edge computing,'' in {\em Proc. WiOpt}, Shanghai, China, May 2018, pp. 1--5.

\bibitem{NOMA}
F. Wang, J. Xu, and Z. Ding, ``Optimized multiuser computation offloading with multi-antenna NOMA,'' in {\em Proc. IEEE GLOBECOM Workshops-NOMAT5G}, Singapore, Dec. 2017, pp. 1--7.

\bibitem{17}
B. McMahan and D. Ramage, ``Federated learning: Collaborative machine learning without centralized training data,'' Tech. Rep., Apr. 2017. [Online]. Available: \url{https://research. googleblog. com/2017/04/federated-learning-collaborative. html}

\bibitem{18}
R. Corless, G. Gonnet, D. Hare, D. Jeffrey, and D. Knuth, ``On the Lambert W function,'' {\it Adv. Comput. Math.}, vol. 5, no. 1, pp. 329--359, Dec. 1996.

\bibitem{19}\label{cvx}
S. Boyd and L. Vandenberghe, {\it Convex Optimization}, Cambridge, U.K.: Cambridge Univ. Press, Mar. 2004.

\bibitem{20}
S. Boyd, ``Ellipsoid method,''  Stanford Univ., Stanford, CA, USA, Tech. Rep., May 2014. [Online]. Available: \url{https://web.stanfordedu/class/ee364b/lectures/ellipsoid method slides.pdf}


\end{thebibliography}
\end{document}